\documentclass[aps,prl,preprint,showpacs,groupedaddress]{revtex4}
\usepackage{graphicx}
\usepackage{amsmath,bm,calc}
\usepackage[psamsfonts]{amssymb}

\begin{document}

%\preprint{}

\title{Rearrangement in folding potentials
with density-dependent nucleon-nucleon interaction}


\author{H. Nakada}
\email{nakada@faculty.chiba-u.jp}
%\thanks{}
\affiliation{Department of Physics, Faculty of Science,
Chiba University, Inage, Chiba 263-8522, Japan}
\author{T. Shinkai}
\affiliation{Graduate School of Science and Technology, 
Chiba University, Inage, Chiba 263-8522, Japan}


\date{\today}

\begin{abstract}
 We discuss optical potentials for the nuclear elastic scattering
 from a variational viewpoint.
 Density-dependence in the effective $N$-$N$ interaction
 leads to density rearrangement terms,
 in addition to the conventional folding term.
 Effects of the rearrangement on the $N$-$A$ optical potential
 are illustrated in the nuclear-matter limit.
 Closely relating to consistency with the saturation,
 the rearrangement appreciably improves
 the isoscalar optical potential depth
 over the previous folding model calculations.
 The rearrangement gives stronger effects as the density grows.
 We also present rearrangement terms
 in the $A$-$A$ double-folding potential.
 Since the rearrangement terms are relevant to the nuclear structure
 but should be handled within the reaction model,
 $N$-$N$ effective interactions applicable both to structure models
 and to the folding model will be desired for unambiguous description
 of the nuclear elastic scattering.
\end{abstract}

%\keywords{heat capacity of nuclei, finite-temperature BCS theory,
% particle-number projection}

\maketitle

%\section{}
%\subsection{}
%\subsubsection{}

% If in two-column mode, this environment will change to single-column
% format so that long equations can be displayed. Use
%\begin{widetext}
%\end{widetext}

The optical potential for elastic scattering
is basic to describing various nuclear reactions~\cite{BD04};
the direct reactions including inelastic scattering,
charge exchange and nucleon transfer reactions,
which are treated in the distorted-wave Born approximation (DWBA),
as well as the compound nuclear reactions.
The folding model has successfully been applied
to obtain the optical potentials,
their real parts in particular,
based on effective $N$-$N$ interactions~\cite{SL79}.

In regard to the effective $N$-$N$ interaction,
density-dependence is required to reproduce the saturation
of nuclear densities and energies,
as qualitatively disclosed by the Brueckner theory~\cite{FW71}.
However, Brueckner's $g$-matrix does not reproduce
the nuclear saturation quantitatively
(as long as we apply it with two-body interaction
within the non-relativistic framework).
In most nuclear structure problems
we use phenomenological effective interactions
in which density-dependence is contained
so as to reproduce the saturation~\cite{VB72}.
Density-dependence has also been introduced
in the effective $N$-$N$ interaction
for the folding model~\cite{DDM3Y}.
Efforts have been made to connect the folding-model interaction
to the saturation~\cite{BDM3Y,KOB94}.
It seems hopeful to unify the folding model
at relatively low incident energies
and nuclear structure models
such as the mean-field approximation~\cite{HL98,Nak03}.
However, there exists inconsistency between them
in treating the density-dependence.
In this article we reconsider the folding model
from a variational viewpoint,
and show where the inconsistency comes from
and how it should be resolved.

We here express an effective $N$-$N$ interaction in the following form:
\begin{equation}
 \hat{v} = \sum_n C_n[\rho]\,\hat{w}_n\,,
\label{DDint}\end{equation}
where $\hat{w}_n$ is a two-body interaction operator,
$C_n[\rho]$ its density-dependent coupling constant.
As in Refs.~\cite{DDM3Y,BDM3Y,KOB94},
the coupling constants are assumed to depend
on the total density $\rho$.
Although there could also be energy-dependence
in the coupling constants,
it does not play an essential role in the arguments below.

We first consider the $N$-$A$ scattering.
The system has $(A+1)$ nucleons, with the total wave-function
$|\Psi_{A+1}\rangle$ under the total Hamiltonian $\hat{H}_{A+1}$.
We decompose $|\Psi_{A+1}\rangle$ as
\begin{equation}
 |\Psi_{A+1}\rangle \approx |\Psi_A\rangle\otimes|\psi_N\rangle\,,
\label{decomp}\end{equation}
where $|\Psi_A\rangle$ is the wave-function of the target nucleus $A$,
and $|\psi_N\rangle$ denotes the wave-function
of the incident and scattered nucleon.
In the folding model, we customarily assume $|\Psi_A\rangle$
to be the ground-state function $|\Psi_A^{(0)}\rangle$,
which satisfies $\hat{H}_A|\Psi_A^{(0)}\rangle
= E_A^{(0)}|\Psi_A^{(0)}\rangle$.
This assumption is practically equivalent
to the frozen density approximation.
With the decomposition (\ref{decomp}),
the Schr\"{o}dinger equation for $\psi_N$
(or the Dirac equation in relativistic approaches,
with appropriate subtraction of the mass)
is derived by the variational equation,
\begin{equation}
 \frac{\delta}{\delta\psi_N^\ast} \left[\langle\Psi_{A+1}|
 (\hat{H}_{A+1}-E) |\Psi_{A+1}\rangle\right]=0\,,
\label{Sch-eq-total}\end{equation}
which can be written as
\begin{equation}
 \hat{h}_N|\psi_N\rangle= E_N |\psi_N\rangle\,;\quad
 \hat{h}_N = \hat{K}_N + \hat{U}\,,~~E_N = E-E_A\,,
\label{Sch-eq-N}\end{equation}
where $E_A=\langle\Psi_A|\hat{H}_A|\Psi_A\rangle\,(\approx E_A^{(0)})$,
$\hat{K}_N$ stands for the kinetic energy operator
and $\hat{U}$ is regarded as the optical potential.
In applying Eq.~(\ref{decomp}) to (\ref{Sch-eq-total}),
one should notice that $\rho_{A+1}$ obtained from $|\Psi_{A+1}\rangle$
should be used in $C_n[\rho]$ of $\hat{H}_{A+1}$,
while $\hat{H}_A$ contains $\rho_A$
which is calculated from $|\Psi_A^{(0)}\rangle$.
Having certain contribution to $\hat{U}$,
$\delta\rho=\rho_{A+1}-\rho_A=\psi_N^\ast \psi_N$
is not negligible though infinitesimal at each position.
We then obtain
\begin{equation}
 \hat{U}|\psi_N\rangle = \sum_n\left\{
 C_n[\rho_A]\cdot\langle\Psi_A|\hat{w}_n\big(|\Psi_A\rangle
 \otimes|\psi_N\rangle\big)
 + \left\langle\frac{\delta C_n}{\delta\rho}\,
   \delta({\textstyle\frac{\mathbf{r}_i+\mathbf{r}_j}{2}-\mathbf{r}_N})
   \,\hat{w}_n\right\rangle_A
 \cdot |\psi_N\rangle \right\}\,.
\label{Utot}\end{equation}
Here $\langle O\rangle_A \equiv\langle\Psi_A|O|\Psi_A\rangle$,
and $\delta(\frac{\mathbf{r}_i+\mathbf{r}_j}{2}-\mathbf{r}_N)$
equates the center-of-mass position of the interacting nucleons in $A$
to the position of the scattered nucleon.
The first term on the rhs in Eq.~(\ref{Utot})
is the usual folding potential.
It is remarked that the density rearrangement term (DRT) comes out
as the second term.
%The DRT is not a dynamical polarization effect,
%since it is not connected to excitation of $A$.
While such rearrangement terms are known
in structure models and in some reaction approaches (\textit{e.g.}
the one within the time-dependent Hartree-Fock approximation),
the DRT has not been taken into account in the folding model.
We call $\hat{U}$ in Eq.~(\ref{Utot})
\textit{renormalized folding potential (RFP)},
because of the correspondence
to the renormalized Brueckner-Hartree-Fock approach~\cite{JLM77}
as discussed later.
The RFP differs from the conventional folding potential
because of the presence of the DRT.
%Physical meaning of the rearrangement term will be argued later.

To illustrate significance of the DRT,
we consider the nucleon scattering on the uniform nuclear matter
with equal numbers of protons and neutrons.
This nuclear-matter folding potential corresponds to depth
of the isoscalar optical potential in the $N$-$A$ scattering.
As far as contribution of the dynamical polarization
is negligibly small,
the RFP is
\begin{equation}
 U(k;k_\mathrm{F}) = \sum_n\left\{
 C_n[\rho]\,\frac{\Omega}{(2\pi)^3}\sum_{\sigma'\tau'}
 \int_{k'\leq k_{\mathrm{F}}}d^3k'\,
 \langle\mathbf{k}\sigma\tau\,\mathbf{k}'\sigma'\tau'|\hat{w}_n
 |\mathbf{k}\sigma\tau\,\mathbf{k}'\sigma'\tau'\rangle
 + \frac{1}{\Omega}\,\frac{\delta C_n[\rho]}{\delta\rho}
 \langle\hat{w}_n\rangle \right\}\,,
\label{U-matter}\end{equation}
where $k_\mathrm{F}$ denotes the Fermi momentum,
and $\Omega$ is the volume of the nuclear matter.
We have $\rho=(2/3\pi^2)k_\mathrm{F}^3$, as usual.
Note that $U$ becomes independent of the spin and isospin component
$\sigma$ and $\tau$ in the symmetric nuclear matter,
as well as of direction of the momentum $\mathbf{k}$.
Obviously $U(k;k_\mathrm{F})$ is connected
to the single-particle (s.p.) energy.
The s.p. energy is defined by
variation of the nuclear matter energy $E$
with respect to the occupation number $n_{\mathbf{k}\sigma\tau}$,
which is unity for $k\leq k_\mathrm{F}$
and vanishes for $k>k_\mathrm{F}$.
The nuclear matter energy is expressed by
\begin{eqnarray}
 E &=& \frac{\Omega}{(2\pi)^3} \sum_{\sigma\tau} \int d^3k\,
 \frac{k^2}{2M}\,n_{\mathbf{k}\sigma\tau} \nonumber\\
 &&+ \frac{1}{2} \sum_n C_n[\rho]\,\frac{\Omega^2}{(2\pi)^6}
 \sum_{\sigma\tau\sigma'\tau'}
 \int d^3k\,d^3k'\,
 \langle\mathbf{k}\sigma\tau\,\mathbf{k}'\sigma'\tau'|\hat{w}_n
 |\mathbf{k}\sigma\tau\,\mathbf{k}'\sigma'\tau'\rangle\,
 n_{\mathbf{k}\sigma\tau}\,n_{\mathbf{k}'\sigma'\tau'}\,,
\end{eqnarray}
and therefore
\begin{equation}
 \varepsilon(k;k_\mathrm{F})
 = \frac{\delta E}{\delta n_{\mathbf{k}\sigma\tau}}
 = \frac{k^2}{2M} + U(k;k_\mathrm{F})\,.
\label{matter-spe}\end{equation}
This $\varepsilon(k;k_\mathrm{F})$ is equivalent to $E_N$
in Eq.~(\ref{Sch-eq-N}) in the scattering problems,
manifesting similarity of the folding model
to the mean-field picture.
Let us now recall the Hugenholtz-van Hove (HvH) theorem~\cite{HvH}.
The saturation point is obtained as the minimum
of $\mathcal{E}\equiv E/A=E/\rho\,\Omega$.
Since the Fermi energy is given by
$\varepsilon_\mathrm{F}\equiv \varepsilon(k_\mathrm{F};k_\mathrm{F})
=(1/\Omega)\,(\partial E/\partial\rho)$,
we obtain
\begin{equation}
 \varepsilon_\mathrm{F}=\mathcal{E}
  +\rho\frac{\partial\mathcal{E}}{\partial\rho}\,.
\end{equation}
This yields $\varepsilon_{\mathrm{F}0}\equiv
\varepsilon(k_{\mathrm{F}0};k_{\mathrm{F}0})=\mathcal{E}_\mathrm{min}$
at the saturation point,
where $k_{\mathrm{F}0}$ represents the Fermi momentum at the saturation
and $\mathcal{E}_\mathrm{min}$ the saturation energy.
We then have~\cite{HM72}
\begin{equation}
 U(k_{\mathrm{F}0};k_{\mathrm{F}0})
  = \mathcal{E}_\mathrm{min}-\frac{k_{\mathrm{F}0}^2}{2M}\,.
\label{U-sat-value}\end{equation}
Since the saturation is directly linked
to the variation of $\mathcal{E}$,
this relation is essential to consistency with the saturation.
Furthermore, the effective mass at the saturation point, $M_0^\ast$,
is given by
\begin{equation}
 \frac{k_{\mathrm{F}0}}{M_0^\ast} \equiv \left.
  \frac{\partial\varepsilon(k;k_{\mathrm{F}0})}{\partial k}
  \right\vert_{k=k_{\mathrm{F}0}}
 = \frac{k_{\mathrm{F}0}}{M} + \left.
  \frac{\partial U(k;k_{\mathrm{F}0})}{\partial k}
  \right\vert_{k=k_{\mathrm{F}0}}\,,
\end{equation}
which derives
\begin{equation}
 \left.\frac{\partial U(k;k_{\mathrm{F}0})}{\partial E_N}
  \right\vert_{k=k_{\mathrm{F}0}}
 = 1-\frac{M_0^\ast}{M}\,.
\label{U-sat-slope}\end{equation}
Here $E_N=\varepsilon(k;k_\mathrm{F})$ is related
to $k$ via Eq.~(\ref{matter-spe}).
Note that, even without energy-dependent coupling constants,
non-locality in the interaction gives rise to
$k$-dependence of $U$, leading to $E_N$-dependence.
For the scattering problems, $E_N$ must be positive.
However, with the extrapolation $k\rightarrow k_{\mathrm{F}0}$
(\textit{i.e.} $E_N\rightarrow\mathcal{E}_\mathrm{min}
\approx -16\,\mathrm{MeV}$),
$U(k;k_{\mathrm{F}0})$ is constrained
by Eqs.~(\ref{U-sat-value},\ref{U-sat-slope}),
as long as it is continuous.

\begin{figure}
\includegraphics[scale=0.85]{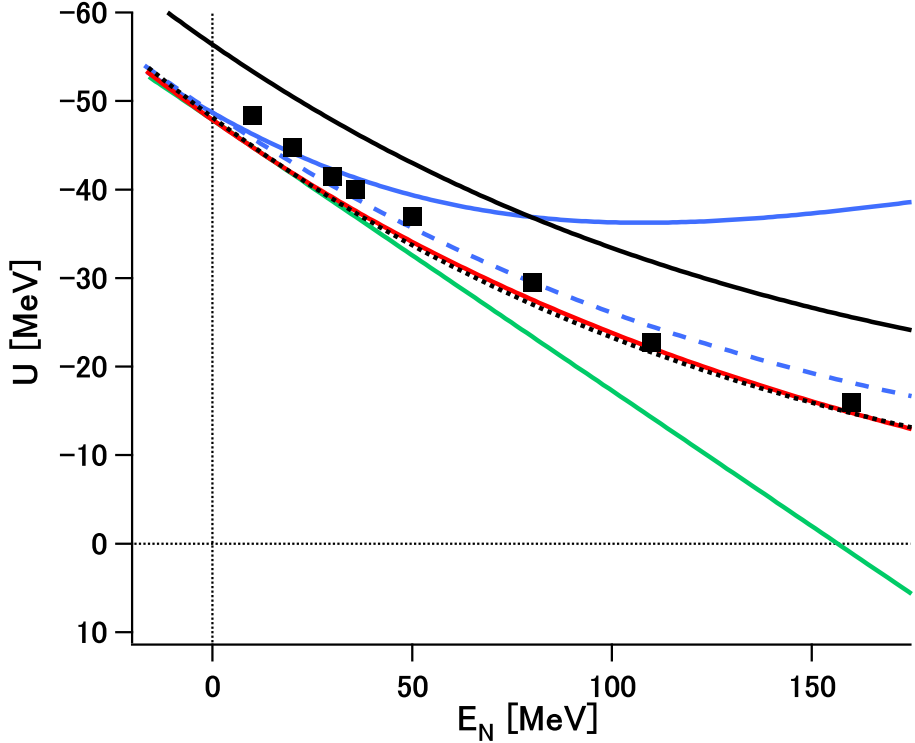}%
\caption{Energy-dependence of the folding potential
 for the nucleon scattering on the nuclear matter,
 at $\rho=0.16\,\mathrm{fm}^{-3}$.
 Results from several $N$-$N$ effective interactions are compared;
 Skyrme SLy4~\cite{SLy} (green line),
 Gogny D1~\cite{D1} (blue dashed line)
 and D1S~\cite{D1S} (blue solid line),
 M3Y-P2~\cite{Nak03} (red line),
 BDM3Y1~\cite{BDM3Y} without the DRT (black solid line)
 and that with the DRT (black dotted line).
 Experimental values are represented by squares,
 which are obtained from depth of the real part
 of the isoscalar optical potential~\cite{BM1}.
 \label{fig:Uk_E}}
\end{figure}

Figure~\ref{fig:Uk_E} shows $U(k;k_\mathrm{F})$ of Eq.~(\ref{U-matter})
as a function of $E_N$ at $\rho=0.16\,\mathrm{fm}^{-3}$.
The calculations are implemented with several interactions
used in the mean-field approaches
as well as with a density-dependent interaction
developed for the folding model, BDM3Y1~\cite{BDM3Y}.
In the mean-field interactions,
a $\rho$-dependent coupling constant is introduced
for a contact term of the interaction.
All of their results meet almost at the same point
by extrapolation to $E_N\approx -16\,\mathrm{MeV}$,
fulfilling the HvH theorem.
They also give slopes close to one another
at $E_N\approx -16\,\mathrm{MeV}$,
corresponding to $M_0^\ast\approx 0.7M$~\cite{Nak03,Lan-Sky}.
However, as $E_N$ increases,
$U(k;k_\mathrm{F})$ behaves differently
among various effective interactions.
This is not surprising because these mean-field interactions
are determined from the structural information,
which is not sensitive to $U(k;k_\mathrm{F})$
far from the saturation point.
The Skyrme interaction~\cite{VB72,SLy} necessarily gives
linear dependence on $E_N$.
While D1S~\cite{D1S} does not give reasonable $E_N$-dependence,
D1~\cite{D1} and M3Y-P2~\cite{Nak03} have $E_N$-dependence
similar to that of the empirical depth of the optical potential.
There is a class of effective interactions in which
the M3Y interaction~\cite{M3Y} is multiplied
by a $\rho$-dependent factor~\cite{DDM3Y,BDM3Y,KOB94}.
We here treat BDM3Y1 as a representative of them.
For BDM3Y1, both the results with and without the DRT
are presented.
We do not use the factor giving additional energy-dependence
(denoted by $g(E_N)$ in Ref.~\cite{BDM3Y}).
Although BDM3Y1 is adjusted so as to reproduce $k_{\mathrm{F}0}$
and $\mathcal{E}_\mathrm{min}$,
the folding potential calculated with BDM3Y1 in Ref.~\cite{BDM3Y}
is contradictory to Eq.~(\ref{U-sat-value}),
because the DRT is discarded.
If the DRT is added,
depth of the folding potential with BDM3Y1
is much closer to the empirical values,
and is almost indistinguishable from that with M3Y-P2.
Thus the RFP improves the optical potential depth
over the previous folding model approach
to a substantial extent, due to the presence of the DRT.
Even the additional $E_N$-dependent factor seems unnecessary
for BDM3Y1 at $E_N\lesssim 200\,\mathrm{MeV}$.
This result demonstrates significance of the constraint
(\ref{U-sat-value}).

\begin{figure}
\includegraphics[scale=0.85]{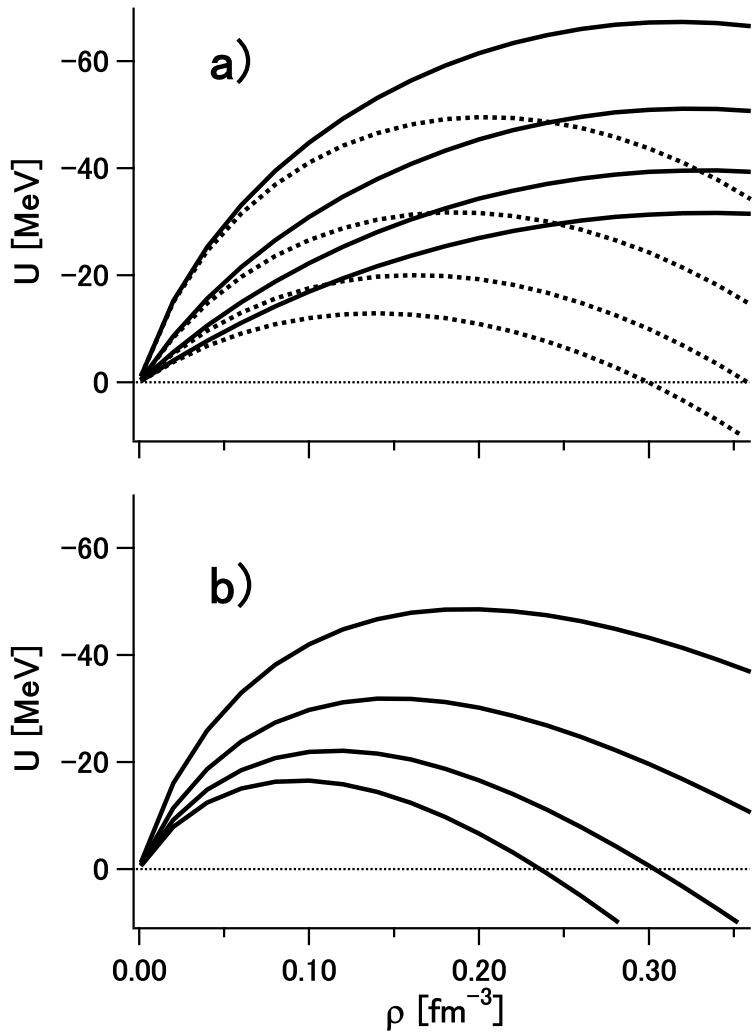}%
\caption{Density-dependence of the folding potential
 for the nucleon scattering on the nuclear matter
 at $E_N=0$, $60$, $120$ and $180$\,MeV
 (from the upper line to the lower).
 a) Results from BDM3Y1, with (dotted line) and without (solid line)
 the DRT.
 b) Results from M3Y-P2,
 for which the DRT is taken into account.
 \label{fig:Uk_rho}}
\end{figure}

In Fig.~\ref{fig:Uk_rho}, $\rho$-dependence of $U(k;k_\mathrm{F})$
is depicted for several $E_N$.
Because of the $E_N$-dependence presented in Fig~\ref{fig:Uk_E},
the folding potential becomes less attractive, or even repulsive,
as $E_N$ increases.
As viewed in the BDM3Y1 results,
effects of the rearrangement are the stronger
for the higher $\rho$.
Therefore, although the rearrangement will not seriously influence
highly peripheral reactions,
it affects the volume integral of the potential.
It will be important to reassess reaction calculations
by applying the RFP.
Whereas BDM3Y1 without the DRT
gives attractive $U(k;k_\mathrm{F})$
even at twice the saturation density,
the RFP comes repulsive at high $\rho$ for relatively high $E_N$.
This $\rho$-dependence is qualitatively similar
to that with M3Y-P2.
Another interesting point suggested by Fig.~\ref{fig:Uk_rho}
will be that the DRT tends to produce or enhance
the so-called wine-bottle-bottom shape in the optical potential
at a certain energy region,
which significantly influences the analyzing power~\cite{RHC92}.

\begin{figure}
\includegraphics[width=15cm]{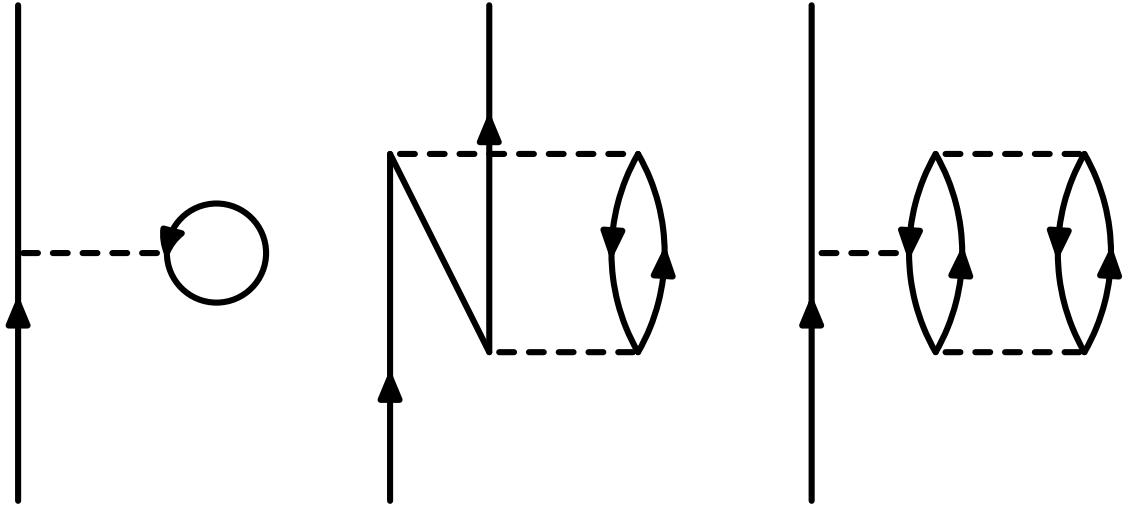}%
\caption{Goldstone diagrams regarding the optical potential.
 Left: 1st-order diagram in the hole-line expansion.
 Middle: 2nd-order diagram due to rearrangement
 for the Pauli exclusion operator.
 Right: 2nd-order diagram due to rearrangement
 for the energy denominator.
 Exchange diagrams are not shown.
 Solid and dashed lines represent the nucleon
 and the $g$-matrix, respectively.
 \label{fig:diag}}
\end{figure}

Whereas we have assumed the interaction form of Eq.~(\ref{DDint}),
similar arguments hold for the microscopic $N$-$N$ interaction
obtained in terms of the $g$-matrix~\cite{JLM77,HM72,JLM76,PHint,ADG00}.
From the $g$-matrix viewpoints,
the density-dependence in the effective interaction
arises from the Pauli principle,
which affects the Pauli exclusion operator
and the self-energy in the energy denominator.
In the $N$-$A$ scattering,
the projectile shifts the energy of the target nucleus
because of the Pauli blocking;
the s.p. state occupied by the projectile is blocked
in addition to those occupied by the nucleons in the target nucleus.
This additional blocking leads to a DRT,
which is equivalent to the DRT in Eq.~(\ref{Utot}).
The density rearrangement in the $g$-matrix was discussed
in Refs.~\cite{HM72,JLM76}.
In terms of the hole-line expansion,
the density rearrangement is primarily represented
by the 2nd-order diagrams (rearrangement
for the Pauli exclusion operator and for the energy denominator)
shown in Fig.~\ref{fig:diag}~\cite{JLM76},
while the conventional folding takes account
only of the 1st-order diagram.
In this regard the RFP is compared
to the renormalized Brueckner-Hartree-Fock approach in Ref.~\cite{JLM77}.
Relevance of the rearrangement to the HvH theorem
was also pointed out in Ref.~\cite{JLM76}.
However, its effects and importance in the optical potential
have not been recognized sufficiently so far.
The present results indicate that the DRT is not negligible.
The DRT does not give all the 2nd order terms
in the hole-line expansion.
It seems that, rather than the order in the hole-line expansion,
the constraint of Eq.~(\ref{U-sat-value}) plays an important role.

We next turn to the $A$-$A$ scattering.
Let us denote the target nucleus and the projectile
by $A_1$ and $A_2$, respectively.
In this case we have the following DRTs in the RFP,
\begin{equation}
 U^\mathrm{rearr} = \sum_n \left\{
 \big\langle\big(C_n[\rho_{A_1}+\rho_{A_2}]-C_n[\rho_{A_1}]\big)\,
 \hat{w}_n\big\rangle_{A_1}
 + \big\langle\big(C_n[\rho_{A_1}+\rho_{A_2}]-C_n[\rho_{A_2}]\big)\,
 \hat{w}_n\big\rangle_{A_2} \right\}\,.
\label{AA-rearr}\end{equation}
The first term on the rhs in Eq.~(\ref{AA-rearr}) is evaluated
only from $|\Psi_{A_1}\rangle$,
with displacement of the $\rho$-dependent coupling constant
in the interaction of Eq.~(\ref{DDint});
\textit{vice versa} for the second term.
As the $A_2$ nucleus approaches $A_1$, the density changes,
causing the energy shift
through the $\rho$-dependence in the interaction.
Note that, in the $\rho_{A_2}\rightarrow 0$ limit,
the first term recovers the DRT in Eq.~(\ref{Utot}).

To discuss the density rearrangement in the $A$-$A$ scattering
from the $g$-matrix viewpoints,
we shall assume that the momentum is approximately
a good quantum number in the s.p. states, for simplicity.
If the $A_1$ nucleus is isolated, nucleons in $A_1$
occupy the s.p. levels of $k\leq k_{\mathrm{F},A_1}$,
and this distribution gives rise to the Pauli blocking effects
among the nucleons in $A_1$.
When the $A_2$ nucleus is present nearby,
the s.p. levels of $|\mathbf{k}-\mathbf{K}|\leq k_{\mathrm{F},A_2}$
are also blocked
($\mathbf{K}$ is the relative momentum of $A_2$ to $A_1$).
This additional blocking affects the $A_1$ energy,
producing a DRT
which corresponds to the first term in Eq.~(\ref{AA-rearr}).
We note that the Pauli blocking due to the two Fermi spheres
in the $A$-$A$ collision was considered in Ref.~\cite{IKF80},
although relation to the folding model was not clarified.
It is often a good approximation
to replace the Pauli blocking effects
by a function of the total density~\cite{SL79}.
Under this approximation,
we go back to the same arguments as those based on Eq.~(\ref{DDint}),
which have yielded the DRTs
in Eqs.~(\ref{Utot},\ref{AA-rearr}).

The DRTs represent energy shift of the nuclei
during the scattering process.
They should be handled within the reaction model,
because they affect the nuclear properties dynamically.
Now the following question is raised:
should the interaction $\hat{v}$ in Eq.~(\ref{AA-rearr}) be
the interaction in the structure model or that in the reaction model?
Conventionally, separability of nuclear reaction problems
from nuclear structure has been postulated,
and different effective interactions
have been used between the nuclear reaction theory
and the nuclear structure theory.
However, there are many cases in which reaction problems
strongly couple to nuclear structure.
Unified description using consistent interactions is desirable
for total understanding of the phenomena.
The above question suggests that the DRTs
are located at an intersection of the structure theory
and the reaction theory,
exemplifying importance of unified description.
%For instance, without the DRTs,
%consistency between the folding model and the saturation is lost,
%as shown above.
%In other words, the occurrence of the DRTs indicates
%that, under the presence of the density-dependent interaction,
%nuclear reactions are not fully separable
%from structure of the nuclei involved.
To avoid ambiguity, consistent effective interactions
should be applied to structure and reaction calculations.
If we restrict ourselves to reactions at relatively low energies,
it might be possible to develop an interaction
applicable both to structure and reaction calculations.
Analysis like that in Fig.~\ref{fig:Uk_E}
may be useful for first selection of effective interactions
for that purpose.
%Figure~\ref{fig:Uk_E} seems to help searching candidates
%(or good starting points) for that purpose;
%D1 and M3Y-P2, which have been tested in some structure calculations,
%are among them.
%If the rearrangement is taken into account,
%BDM3Y1 could be another candidate.

In low-energy nuclear reactions,
$\rho$-dependence of the effective interaction is important,
and could somewhat be correlated with the saturation.
In contrast, $\rho$-dependence is not so significant
in high-energy reactions.
If the incident energy is sufficiently high,
the impulse approximation works well.
In such energy regime, the phenomenological effective interactions
employed in Figs.~\ref{fig:Uk_E} and \ref{fig:Uk_rho}
will lose their validity.
Furthermore, effects of the density rearrangement
are expected to be weak.
It will be interesting to estimate quantitatively
by the $g$-matrix approach
how large the DRTs are at varying $E_N$.

Many-body forces have similar effects
to the density-dependent two-body interactions~\cite{VB72}.
When a many-body force is introduced
instead of the density-dependent interaction,
we have DRTs again.
For instance, in the $N$-$A$ scattering with a three-body force,
the interaction among the projectile
and two nucleons in the target nucleus gives rise to
energy shift of the target nucleus in an effective manner,
which produces a DRT as in Eq.~(\ref{Utot}).

Effective interactions for inelastic scattering
will also be subject to the density rearrangement effects.
Because the density-derivative of the optical potential is relevant
to the inelastic scattering to collective states
within the DWBA framework~\cite{Gle83},
terms up to $\delta^2 C_n[\rho]/\delta^2\rho$ come into the issue
of the $N$-$A$ process.
This coincides the residual interaction
in the random-phase approximation~\cite{BT75}.
Details will be discussed in a future publication.

The folding model using density-dependent interactions
has been successful in describing nuclear reactions
to a certain degree, without the DRTs.
There could be several reasons for it.
First, some reactions may be insensitive
to the rearrangement effects,
because, \textit{e.g.}, they are highly peripheral.
Second, some of the density-dependent interactions
that have been used in the folding model
were fitted to the reaction data.
This may mask some rearrangement effects.
%Relatedly, there might remain a room to reconsider
%validity of the frozen density approximation.
%A possibility of accidental cancellation
%between the rearrangement and effects of the dynamical polarization
%is not fully excluded.
Furthermore, the dynamical polarization effects should be treated
properly, for full understanding of the optical potential.
%including the absorption effects that induce the imaginary part
%of the potential.
Still it will be important to reinvestigate individual reactions
by taking the DRTs into account.
As stated already, this may open a way to unified description
of the nuclear elastic scattering and the nuclear structure.

In summary,
we have discussed the $N$-$A$ optical potential
based on a variational equation.
Density-dependence in the effective $N$-$N$ interaction
gives a density rearrangement term,
which has been discarded in the conventional folding-model calculations.
We call the optical potential with the rearrangement term
\textit{renormalized folding potential}.
If the rearrangement term is taken into account,
depth of the $N$-$A$ folding potential is in better agreement
with the experimental data
than in the conventional folding model approach.
We emphasize significance of a constraint
due to the Hugenholtz-van Hove theorem,
which assures consistency with the saturation.
Effects of the rearrangement become stronger as the density increases.
Rearrangement terms in the $A$-$A$ scattering are also discussed.
The rearrangement terms seem to be located at an intersection
of the nuclear structure models and the folding model.
It is desired to explore effective $N$-$N$ interactions
applicable to both of these models in a consistent manner,
for unambiguous description
of the low-energy nuclear elastic scattering.

\begin{acknowledgments}
 Discussions with K. Takayanagi and Y. Sakuragi
 are gratefully acknowledged.
 The present work is financially supported, in part,
 as Grant-in-Aid for Scientific Research (B), No.~15340070,
 by the MEXT, Japan.
% A part of numerical calculations are performed on HITAC SR11000
% at Institute of Media and Information Technology, Chiba University.
\end{acknowledgments}


\end{document}